# Universal orbital-coupling rules for hydrogen–defect interactions in bcc metals

Qianxi Zhu[1]†, Hao-Xuan Huang[2,3]†, Zichen Zhang[1]†, Hong-Bo Zhou[2,3]*, Wang Gao[1]*

[1] Key Laboratory of Automobile Materials, Ministry of Education, Department of Materials Science and Engineering, Jilin University, Changchun 130022, China.
[2] Department of Physics, Beihang University, Beijing 100191, China.
[3] Beijing Key Laboratory of Fusion Reactor Materials and Components, Beihang University, Beijing 100191, China.

*Corresponding author(s). E-mail(s): hbzhou@buaa.edu.cn (H.-B. Zhou); wgao@jlu.edu.cn (W. Gao).

†These authors contributed equally to this work.

**Abstract**

Hydrogen–defect interactions control the performance of body-centered-cubic (bcc) metals. However, the complex variations among defects lead to various empirical models that lack transferability across defects. Here, we propose an orbital-coupling model, built on coordination number and interatomic distance, that quantifies the H solution energetics across nanovoids, vacancy loops and grain boundaries in bcc metals, and even predicts the potential-energy surfaces of H at nanovoids. Our model reveals an unconventional s–d coupling rule in the confined environment of defects: H–metal interactions exhibit a unique coordination-dependent law, whereas H–H interactions, deviating from the usually speculated s–s coupling, acquire the distance-decay law of H–metal coupling. This unusual rule proves essential to reproduce the experimentally observed bimodal profile of H desorption. Our electronic-structure-origin, unified model is thus crucial to understanding the nature of chemical bonds under constraint and engineering the H-tolerant materials.

## INTRODUCTION

Hydrogen–defect (H–defect) interactions, which determine H solution in metals, raises pervasive scientific and engineering concerns spanning nuclear-fusion plasma-facing components (*1*–*4*), green-H energy storage (*5*), heterogeneous catalysis (*6*), and H-induced material degradation (*7*–*10*). The low bulk solubility of H in body-centered-cubic (bcc) metals makes the lattice defects, including nanovoids, dislocation loops, and grain boundaries (GBs), the predominant trapping sites for H (*11*–*14*). H–defect interactions, including H–metal (H–M) and H–H interactions at defects, are thus of central importance to understanding the damage mechanism of H-related phenomena and designing high-performance materials for H-applications. Consequently, a long-sought goal has been to build a predictive model of H–defect interactions with a general physical picture across diverse defects.

Among the various defect types, nanovoids have drawn particularly intensive research interest, because of their large H-trapping capacity and their direct relevance to H-bubble nucleation and irradiation-induced void swelling (*15*–*20*). Early effective-medium-theory calculations demonstrated that vacancies furnish energetically favorable local electron-density environments for H accommodation (*21*–*23*), with H atoms preferentially adsorbing onto nanovoid inner surfaces. Subsequent first-principles investigations independently turned to the configuration-dependent site-classification strategies for H solution. Hou *et al.* constructed a predictive empirical model of H–defect interactions by sorting H adsorption sites on nanovoid inner surfaces of tungsten (W) into discrete energy levels based on local atomic configurations (*24*), which reasonably reproduced experimental thermal-desorption-spectroscopy (TDS) measurements when coupled with object kinetic Monte-Carlo (OKMC) simulations. Notably, their empirical model suggested the H–H interactions at nanovoids to decay with the interatomic distance in the power law of five, although this decay behavior would largely eliminate the H–H repulsion in the nanovoids as small as a single vacancy with 2–3 H atoms per vacancy. On this basis, Ma *et al.* further refined the framework by more explicitly incorporating the coordination number effects (*25*), extending its applicability to larger vacancy-clusters (VCs) of W and yielding enhanced TDS-prediction accuracy. However, such empirical models, building upon site-specific geometric categorizations and heavily parameterized against large first-principles datasets, suffer from intrinsic transferability limitations. The descriptors optimized for nanovoid inner surfaces fail to describe the other regions of the potential-energy surfaces like H behavior at void cores, subsurface zones or metastable sites, much less large-scale defects such as dislocation loops and GBs. Previous studies mainly attributed the site-dependent differences of H solution at dislocation loops of iron (Fe) and W to the strain fields, with tensile regions attracting H and compressive regions repelling it (*26, 27*). For H solution at GBs, the alternative methodologies including available-volume evaluation and polyhedral-deformation models have been proposed for H–defect interactions in different metals (*28–30*), yet these are likewise restricted to particular structural motifs and demand manual differentiation across diverse GB configurations and polyhedral units. Such stress arguments at dislocation loops and fragmented treatment at GBs again point to the essential bottleneck in the field: the lack of a unified framework from a more fundamental level—electronic structures—that captures the core physics governing H–defect interactions ranging from nanovoids, dislocation loops, GBs to other defect forms alike.

We have successfully developed an orbital-coupling model for determining the solution energetics of H spanning nanovoids, vacancy loops and GBs. Our model reveals an unconventional s–d coupling rule at defect-confined environments in bcc metals, in which H–M interactions exhibit a unique coordination-dependent law while H–H interactions follow the distance-dependent law of H–M interactions. Furthermore, our model is applicable to diverse metal-defect systems, capturing the potential-energy surfaces of H in nanovoids and reproducing the experimental TDS of H in bcc metals. Our framework establishes a theoretical basis for previous semi-empirical models, a unified physical picture for understanding the nature of chemical bonds under constraint and a practical tool for engineering the H-related properties of metals.

## RESULTS

### Orbital-coupling model of H–defect interactions

H–defect interactions in bcc metals are determined by the coupling between the s-state of H and the d-bands and s-bands of metals (*31, 32*). In the tight-binding (TB) models, the delocalized and half-filled s-bands generate the constant s–s coupling across transition-metals (TMs), while the d-bands determine the trend of interatomic bonding from one metal to another (*31, 33*). The unchanged s-state of H will also produce the constant s–s coupling of H–M interactions at different defects, therefore, we focus on the s–d coupling between H s-state and TM d-bands. A H atom is surrounded by *n* neighboring metal atoms, and each metal atom has a specific coordination number ($CN_i$). In the TB models, the d–d coupling strength of a given atom with its first shell neighbors in metals is determined by the d-band width ($W_d$) of this atom, which varies with the *CN* of the atom (*34, 35*), as $W_d \propto \sqrt{CN} \cdot V_{dd}$, where $V_{dd}$ is the interatomic d–d coupling matrix (*36*). This concept also holds for any TM atom around defects. Based on the Muffin-Tin-Orbital theory (*31*), the coupling strength between the s-state of H and the d-states of TM atoms around defects, is proportional to the interatomic band width, as $W_{sd} \propto \sum_{i=1}^{n} W_{s,H} \cdot W_{d,i}$, mainly reflecting the contribution of the $W_{d,i}$ of TM atoms (*37*). Notably, the metal atoms around defects, from one to another, exhibit the variable *CN*. Therefore, it is crucial to incorporate this *CN* dependence of $W_{d,i}$ for metal atoms, leading to:

$$W_{\mathrm{sd}} \propto \sum_{i=1}^{n} V_{\mathrm{ss}} \cdot \sqrt{CN_i} \cdot V_{\mathrm{dd},i} \propto \sum_{i=1}^{n} \sqrt{CN_i} \cdot V_{\mathrm{sd},i} \propto \sum_{i=1}^{n} \sqrt{CN_i} \cdot \frac{r_s r_{d,i}^{3/2}}{d_i^{7/2}} \quad (1)$$

where $V_{\mathrm{sd}}$ is the interatomic coupling matrix between the s-state of H and d-states of TMs (*31*, *38*), $r_{\mathrm{d}}$ is the d-state radius of TM atoms, $r_{\mathrm{s}}$ is the s-state radius of H and $d_i$ is the interatomic distance. It is reasonable to neglect the change of $r_{\mathrm{d}}$ for a given metal caused by the local coordination environment according to the TB model. Recalling that the $r_{\mathrm{s}}$ of H is also unchanged across the different sites of defects, we obtain $W_{\mathrm{sd}} \propto \sum_{i=1}^{n} \sqrt{CN_i}/d_i^{7/2}$. According to the TB model, the interatomic bonding can be divided into the contributions of attractive and repulsive terms, corresponding to bonding ($E_{\mathrm{bond}}$) and anti-bonding ($E_{\mathrm{anti-bond}}$) states, which are determined by the coupling matrix and its square, respectively (*31*, *34*). Thus, for the interatomic bonding dominated by the attractive term, $E_{\mathrm{sol,H}}$ can be described by the descriptor $D_{\mathrm{at}}$ as

$$E_{\mathrm{sol,H}} \propto E_{\mathrm{bond}} \propto D_{\mathrm{at}} = \sum_{i=1}^{n} \frac{CN_i^{1/2}}{d_i^{7/2}} \quad (2)$$

Similarly, for the interatomic bonding dominated by the repulsive term, the relationship between $E_{\mathrm{sol,H}}$ and $D_{\mathrm{re}}$ is given as

$$E_{\mathrm{sol,H}} \propto E_{\mathrm{anti-bond}} \propto D_{\mathrm{re}} = \sum_{i=1}^{n} \frac{CN_i}{d_i^{7}} \quad (3)$$

**Solution of a single H atom at VCs**

Taking W as the model system, we now apply our model to study the energetic characteristics of H at nanovoids. The VCs with different sizes ($V_n$) are considered, including the most stable $V_1$, $V_2$, $V_3$, $V_4$, $V_6$, $V_9$ and $V_{15}$ structures (*16*, *20*), along with the meta-stable but highly symmetric $V_2^{2nn}$ and $V_9^{meta}$ structures (fig. S1). In addition to the inner surface sites that have been the focus of the previous studies (*24*, *25*), we examine the potential sites at the center of VCs and in the subsurface regions. The cutoff radius used to determine the number and *CN* of H-neighboring metal atoms at each defect site are detailed in note S1.

We first investigate the stable solution sites of H in VCs, and then use fully fixed calculations to analyze the solution of H in the specific region (unstable region) of VCs (all H and metal atoms remained fixed). We find that the interactions of H with neighboring W atoms in VCs can be quantified by the attractive term of the H–M orbital coupling, denoted as $D_{\mathrm{at}} = \sum_{i=1}^{n} \frac{CN_i^{1/2}}{d_i^{7/2}}$ (see Eq. 2). Generally, the stable sites of H are primarily located on the inner surface and subsurface layer of VCs, with the corresponding $E_{\mathrm{sol,H}}$ exhibiting a positive linear relationship with our descriptor $D_{\mathrm{at}}$ (Fig. 1A and fig. S2). In contrast, H can be also stable near the centers of the $V_4$ and $V_9^{meta}$, generating a negative linear relationship with $D_{\mathrm{at}}$ until the inflection point (Fig. 1B and fig. S2B). To further probe the H solution at VC centers—most of which are unstable H sites, we examine additional sites near the centers of more VCs using fixed calculations. Notably, the resulting $E_{\mathrm{sol,H}}$ exhibits a negative correlation with $D_{\mathrm{at}}$ (Fig. 1C and fig. S3), the same as found for the stable sites near the VC centers of $V_4$ and $V_9^{meta}$. Apparently, $E_{\mathrm{sol,H}}$ is correlated with our descriptor $D_{\mathrm{at}}$ in broken-line relationship for each VC, showing the opposite trends at different regions of VCs.

We now turn to understand the origin of the broken-line relationship between $D_{\mathrm{at}}$ and $E_{\mathrm{sol,H}}$. Notably, this relationship reflects the physical processes of H sites transitioning from the VC centers to the inner surface sites, and subsequently to the subsurface regions. The determining factor is the competition between the attractive and repulsive H–M interactions, which depends on the H–M distance, the number and *CN* of neighboring W atoms around H (Eq. 2). The region on the left side of the inflection point represents the transition of H from the VC centers to the inner surfaces. This process is driven by the H–M attractive interaction: the decrease of H–M distance generates the decrease of $E_{\mathrm{sol,H}}$ and the increase of descriptor $D_{\mathrm{at}}$, leading to a negative correlation between $D_{\mathrm{at}}$ and $E_{\mathrm{sol,H}}$. In contrast, the data points on the right of the inflection point reflect the continuous transition of H from the inner surface to the subsurface layer sites. During this process, the repulsive H–M interactions play an increasingly important role, increasing $E_{\mathrm{sol,H}}$. Meanwhile, the number and *CN* of neighboring W atoms around H increase, resulting in an increase of $D_{\mathrm{at}}$. Accordingly, a positive correlation between $D_{\mathrm{at}}$ and $E_{\mathrm{sol,H}}$ appears.

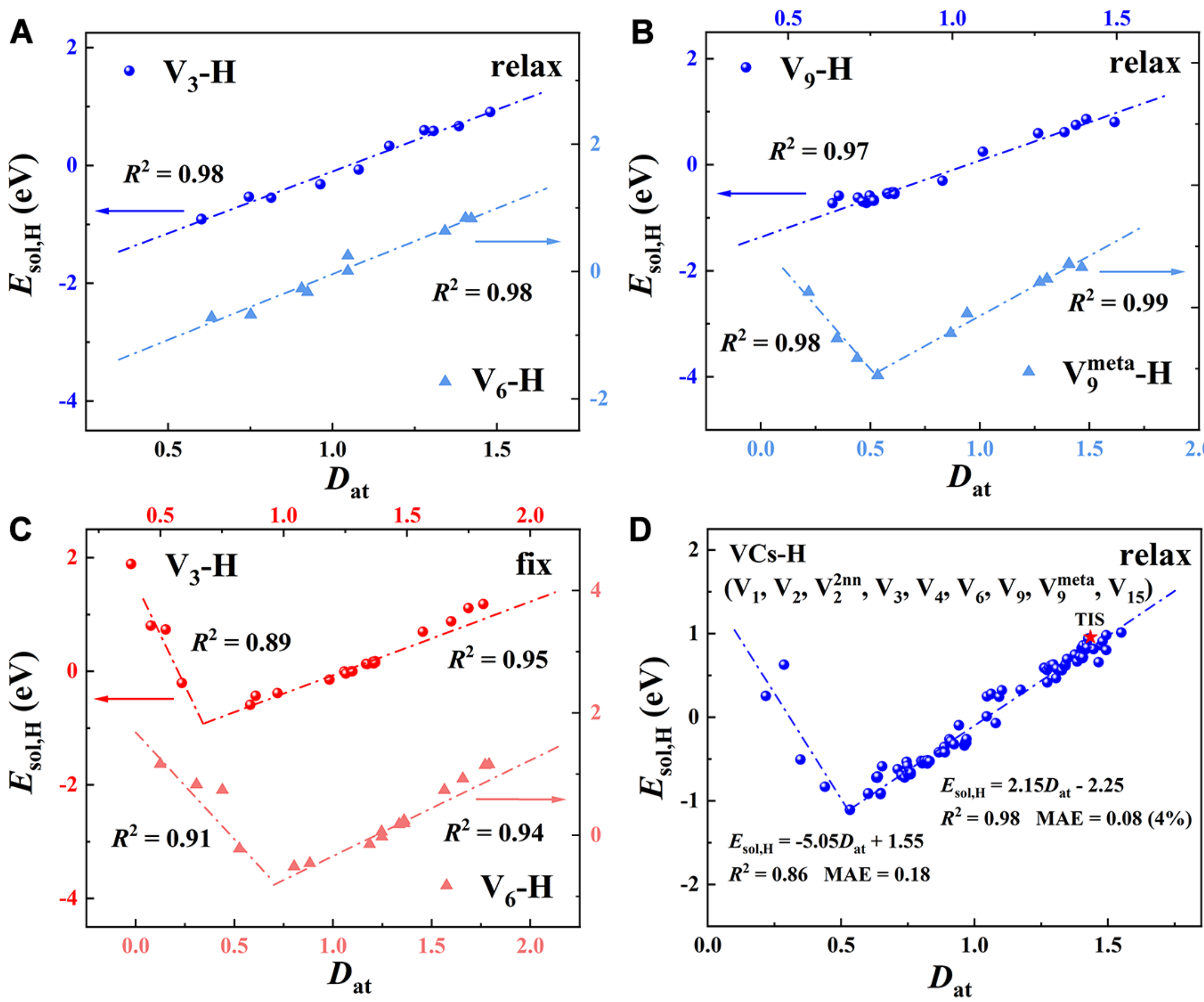

**Fig. 1. The solution energy of a single H ($E_{\mathrm{sol,H}}$) in W VCs.** (**A** and **B**) The $E_{\mathrm{sol,H}}$ under relaxed calculations versus the descriptor $D_{\mathrm{at}}$ in (**A**) $V_3$ and $V_6$ and (**B**) $V_9$ and $V_9^{\mathrm{meta}}$. (**C**) The $E_{\mathrm{sol,H}}$ under fully fixed calculations versus the descriptor $D_{\mathrm{at}}$ in $V_3$ and $V_6$. (**D**) The $E_{\mathrm{sol,H}}$ under relaxed calculations in all studied W VCs versus the descriptor $D_{\mathrm{at}}$.

Notably, our descriptor is capable of quantifying $E_{\mathrm{sol,H}}$ across different size VCs in a broken-line relationship (Fig. 1D). The combination of the H–M distance, the number and *CN* of neighboring W atoms around H clearly provides an effective way to determine the size-dependent effects of VCs. Our scheme also demonstrates that, over a wide range of VC sizes, the $E_{\mathrm{sol,H}}$ in W VCs strongly depends on the d-band width of W atoms around VCs and can be quantified by the attractive term of s–d orbital coupling between H and neighboring W atoms. From a broader perspective, the overall H solution curve in VCs predicted by our descriptor $D_{\mathrm{at}}$ aligns well with the relationship between local electron density and $E_{\mathrm{sol,H}}$ in the effective medium theory (*21*, *23*), suggesting that the descriptor $D_{\mathrm{at}}$ also reflects the local electron density at VCs. Unlike effective-medium theory, however, our descriptor $D_{\mathrm{at}}$ is a purely geometric-dependent quantity, provides an easily accessible and widely applicable quantitative relationship for H in VCs without requiring electron-density calculations. Thus, the descriptor $D_{\mathrm{at}}$ offers a simple, physically grounded and universal prediction of $E_{\mathrm{sol,H}}$ in VCs.

We further explore the potential-energy surfaces of H in W VCs using our descriptor $D_{\mathrm{at}}$. Taking the $V_6$ structure as a representative example (Fig. 2A), we perform fixed calculations to scan $E_{\mathrm{sol,H}}$ within the four-vacancy plane to fit the density functional theory (DFT) potential-energy surface of H solution (Fig. 2B), and then construct the predicted potential-energy surface based on our model predictions (Fig. 2C). Notably, the predicted potential-energy surface is in good agreement with the DFT-calculated one, accurately capturing the overall shape, energy peaks, and continuous energy variations across both stable and unstable sites in the center, inner surface and subsurface regions of $V_6$. Previous studies have primarily focused on identifying energetically favorable H sites at the inner surface of VCs (*20*, *24*, *25*), while neglecting the overall H potential-energy surfaces of VCs, resulting in an incomplete understanding of the energy landscape. In contrast, our model effectively captures the H potential-energy surfaces in W VCs and establishes a comprehensive quantitative structure–property relationship between local geometric structures and $E_{\mathrm{sol,H}}$, which provides a powerful tool for predicting the energetics of H diffusion and aggregation in VCs. Furthermore, it offers a transferable, physically motivated descriptor—based on simple interatomic distances—for developing interatomic potentials for VC–H interactions.

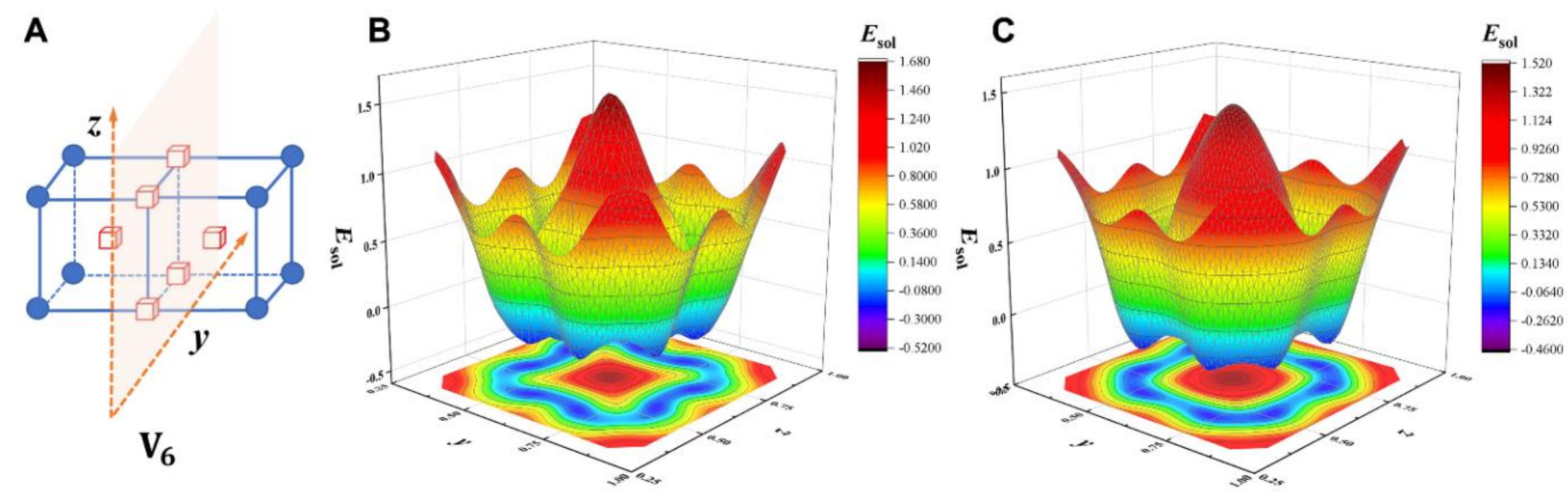

**Fig. 2. Potential-energy surfaces of H solution in the four-vacancy plane of the $V_6$ structure in W.** (**A**) Schematic representation of the selected four-vacancy plane in $V_6$. (**B** and **C**) The ones obtained from (**B**) DFT calculations and derived from (**C**) our model predictions.

**Multi-H in W VCs**

In reality with a certain H concentration, both H–M and H–H interactions are crucial to the H solution in W VCs. To investigate the H–H interactions in W VCs, we construct a series of 2H configurations on the inner surface of VCs with varying H–H distances using DFT calculations. In addition, we fine-tune the graph atomic cluster expansion (GRACE) potential for H–W systems using our calculated DFT data (*39*, *40*), which is capable of reproducing the H solution energetics in VCs with DFT-level accuracy (fig. S4). On this basis, we perform the extensive calculations of the H–H binding energies in VCs over a wide range of H–H distances ($d_{\mathrm{H-H}}$). For the relaxed 2H configurations, we find that $E_{\mathrm{sol,2H}}$ can be effectively quantified by summing the descriptors $D_{\mathrm{at}}$ of two H-sites (as $D_{\mathrm{at,2H}}$) for $V_{n\geq 2}$ (Fig. 3A), because the H–H pairs are spontaneously adjusted to avoid the H–H repulsion.

In contrast to the relaxed systems, the fixed 2H calculations exhibit different energy trends: the H–H interactions contribute significantly to the total solution energy (Fig. 3B), when the H atoms are confined at short distances (corresponding to a high H concentration in VCs). We find that, under interatomic constraints, the H–H interactions exhibit a distance-dependent relationship similar to the s–d coupling of H–M in VCs in a $1/d_{\mathrm{H-H}}^{7/2}$ dependence (see the comparison in Fig. 3, B to D). This reflects a fundamentally different physical picture compared with the literature results. On the one hand, the H–H interactions in VCs are likely speculated to comply with the usual s–s coupling in a $1/d_{\mathrm{H-H}}^{2}$ law. On the other hand, Hou *et al.* employed numerical fitting to determine the H–H interactions in VCs and proposed a $1/d_{\mathrm{H-H}}^{5}$ distance dependence (*24*). It is noteworthy that their fitting was based on only the configurations with the H–H distance longer than ~1.5 Å. However, for the shorter H–H distances (down to 0.9 Å where the H–H interactions are significant), their fitting results of $1/d_{\mathrm{H-H}}^{5}$ exhibit significant deviations. Importantly, the power law, $1/d_{\mathrm{H-H}}^{5}$, indicates the H–H interactions in VCs following the scaling law characteristic of d–d orbital coupling (*38*), which is physically implausible. In the confined environment of VCs, the H–H interactions are constrained and regulated by the s–d coupling of H–M interactions, causing the H–H interactions to follow a $1/d_{\mathrm{H-H}}^{7/2}$ law. Building on the interatomic coupling rules, we establish a comprehensive quantitative framework that incorporates both H–M and H–H interactions within the confined defect environments, enabling the quantification of multi-H solution in W VCs.

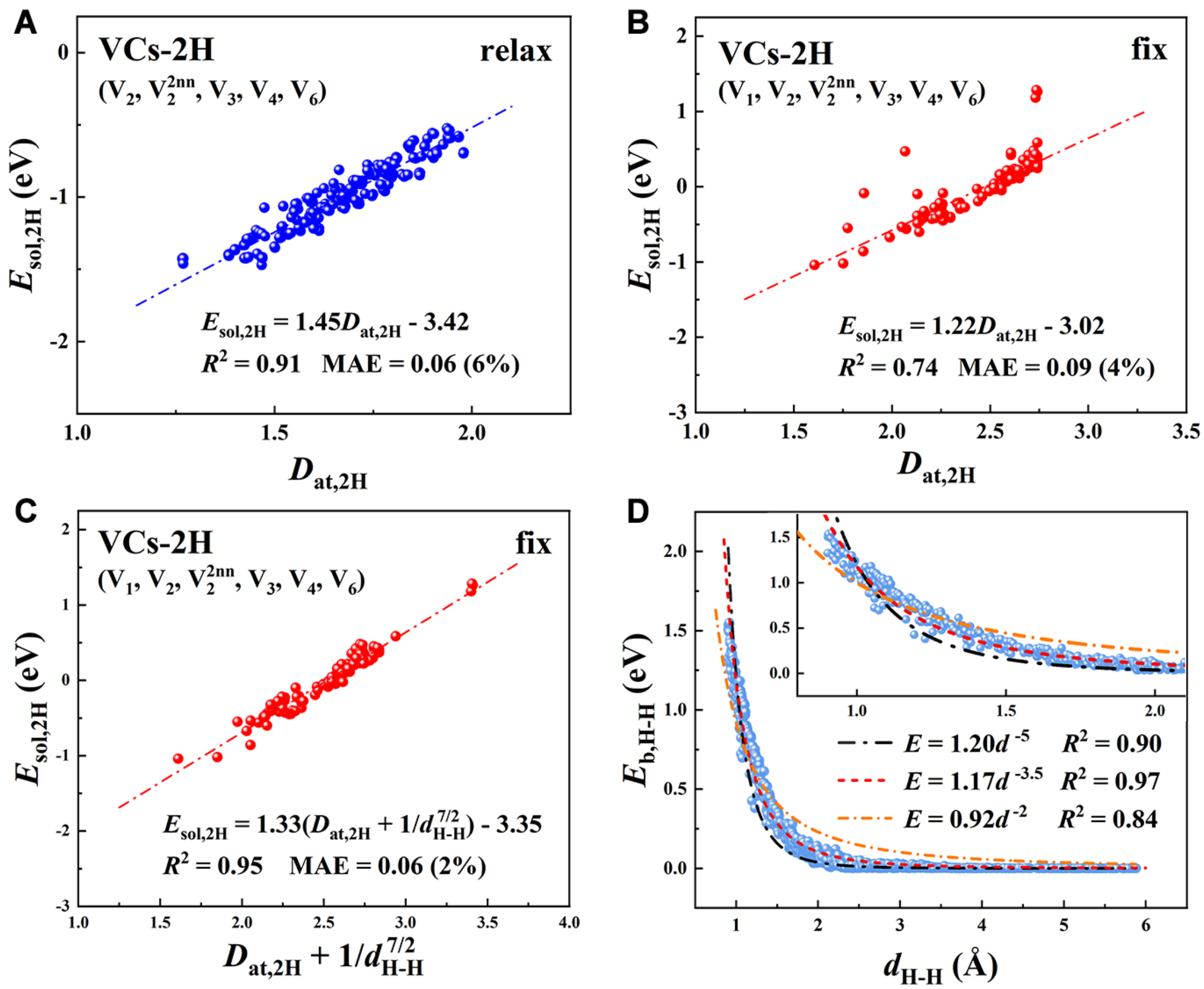

**Fig. 3. The solution energy of two H atoms ($E_{\mathrm{sol,2H}}$) in W VCs along with the corresponding H–H binding energy ($E_{\mathrm{b,H-H}}$).** (**A**) Relaxed $E_{\mathrm{sol,2H}}$ against the H–M interaction descriptor. (**B**) Fully fixed $E_{\mathrm{sol,2H}}$ against the H–M interaction descriptor. (**C**) Fully fixed $E_{\mathrm{sol,2H}}$ against the sum of the H–M and H–H interaction descriptors. (**D**) Power-law relationships between the $E_{\mathrm{b,H-H}}$ and H–H distance.

## H at W vacancy loops

Under irradiation, supersaturated vacancies can migrate and aggregate, eventually collapsing into vacancy-type dislocation loops (*26*, *41*, *42*). To understand the trapping behavior of H at vacancy-type loops, we construct a 1/2 <111> vacancy loop in W. Since dislocation-loops generate long-range stress fields that require large-scale atomistic configurations beyond the scale of direct DFT calculations, we also perform the simulations using the GRACE potential, which was fine-tuned with our DFT data. Our calculations reveal that H exhibits favorable solution at sites near the vacancy-core region of the vacancy loop, in contrast to the unfavorable solution at the outer edge of the loop structure (Fig. 4A). This trend is consistent with the literature findings in the vacancy loop of α-Fe (*26*), reflecting the tensile stress at the core region and the compressive stress at the outer edge. The interatomic bonding of these two different regions thus exhibits the distinct localization characteristics (see note S1 for details): the smaller core region exhibits the more localized stress, relative to the outer edge.

By accounting for the different localization nature of the core region and the outer edge, we find that $E_{\mathrm{sol,H}}$ at the different regions of the vacancy loop exhibits a universal rule that can be well determined by the s–d coupling of H–M interactions (Fig. 4B). Such a simple s–d coupling rule for the H–M interactions at vacancy loops is particularly encouraging, since the significant stress and severe lattice distortion around vacancy loops generally generate a speculation of a more complex structure–property relationship. Recalling that the stress acted as a core argument to account for the H solution at dislocation loops (*26*, *27*, *41*, *42*), our descriptor provides a more fundamental perspective, electronic- and geometric-structure levels, to capture the stress effects on H solution energies. Notably, our model demonstrates that the H–M orbital coupling complies with the same mechanism at vacancy loops and VCs, revealing a consistent physical picture of H solution at vacancy-type defects.

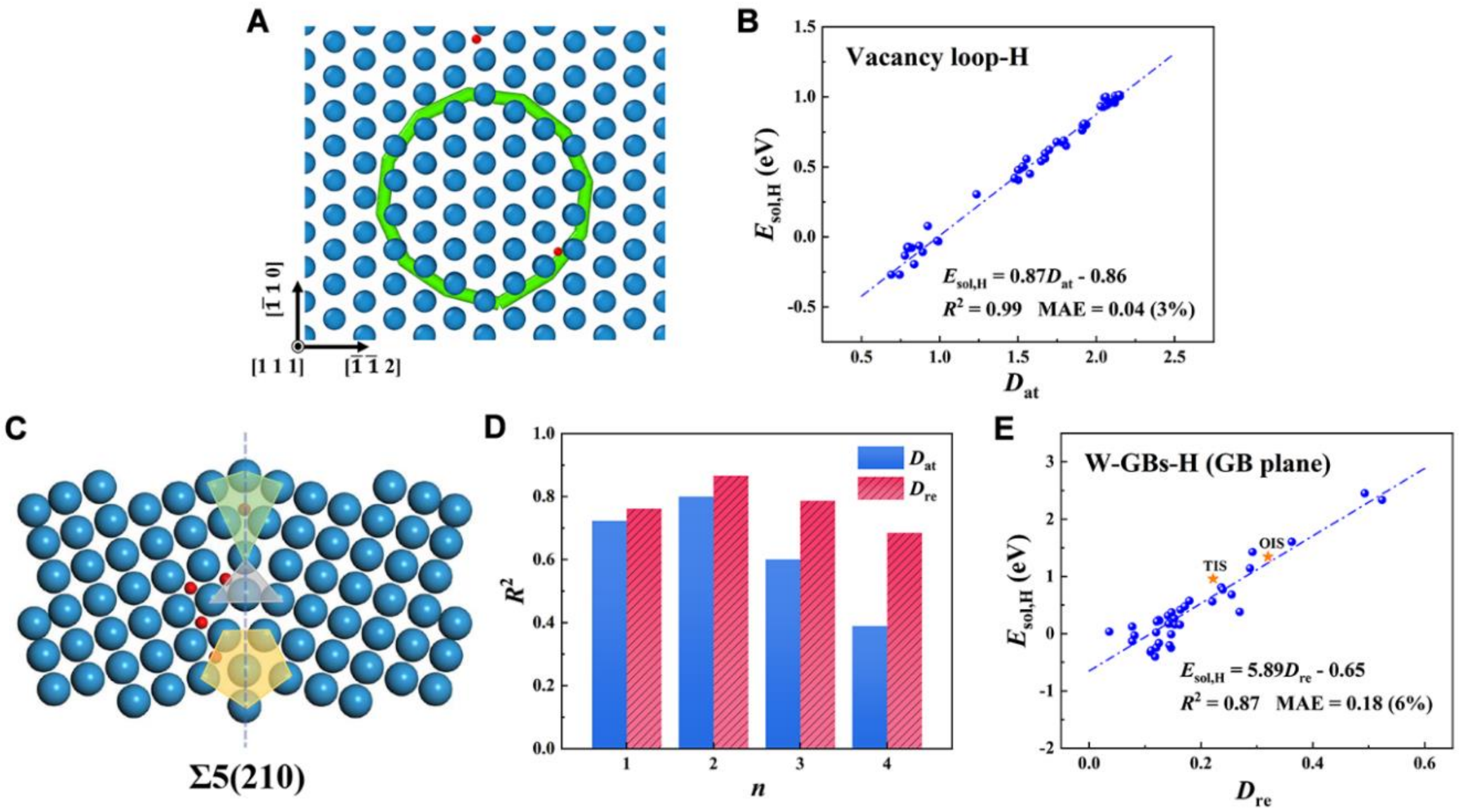

**Fig. 4. The H solution energy ($E_{\mathrm{sol,H}}$) at the vacancy loop and GBs of W.** (**A**) Schematic representation of the 1/2 <111> vacancy-type dislocation loop. (**B**) The $E_{\mathrm{sol,H}}$ of H near the vacancy loop of W versus the descriptor $D_{\mathrm{at}}$. (**C**) Schematic representation of the Σ5(210) GB of W. The interstitial sites at the GB planes are represented by the polyhedral packing units. (**D**) Coefficient of determination ($R^2$) for the linear relationship of the $E_{\mathrm{sol,H}}$ at W GB planes with the attractive and repulsive terms of orbital-coupling with the *n* nearest-neighboring W atoms. (**E**) The $E_{\mathrm{sol,H}}$ of H at W GB planes with our descriptor $D_{\mathrm{re}}$.

### H at W GBs

Now, we investigate the solution of H at W GBs. We construct 12 symmetric tilt GB structures with different plane orientations, focusing on the polyhedral interstitial sites at GB planes (*29*, *30*, *43*), as well as the tetrahedral interstitial sites (TISs) and octahedral interstitial sites (OISs) besides the GB planes (*44*) (Fig. 4C and fig. S5). We find that the H at GBs exhibits more localized interactions with neighboring W atoms. By testing the contribution of the *n* nearest-neighboring W atoms to the H solution, we observe that the optimal linear regression accuracy is achieved when considering the two nearest-neighboring W atoms (Fig. 4D), and thus adopt $n = 2$ for H at GBs in this work. This localized effect indicates the repulsive nature of H–M interactions at GBs. Meanwhile, the repulsive terms of H–M orbital coupling consistently exhibits higher accuracy compared to the attractive terms, further confirming that the repulsive term contributes more prominently to the solution of H at GBs. This stems from the strict geometric constraints imposed by the interstitial space of GBs (*28–30*), which limits the potential distance between H and W atoms. Accordingly, our descriptor for the $E_{\mathrm{sol,H}}$ at GBs is $D_{\mathrm{re}} = \sum_{i=1}^{2} \frac{CN_i}{d_i^7}$.

Our descriptor $D_{\mathrm{re}}$ effectively quantifies the $E_{\mathrm{sol,H}}$ at W GBs (Fig. 4E and fig. S6) and also accurately identifies the solution order of H in TISs and OISs of W bulk. Notably, a pioneering study evaluated $E_{\mathrm{sol,H}}$ at GBs based on the volumetric deformation of polyhedral structural units (*30*, *44*), which required distinguishing different structural units. Instead, our model provides a unified quantification across GB interstitial sites without the need to distinguish GB structures or polyhedron types, which is more convenient for practical applications. Our model also theoretically explains why there is a close relationship between the available volume of interstitial sites and $E_{\mathrm{sol,H}}$ at GBs (*29*): the naturally introduced $d^{7/2}$ term of s–d coupling is close to the power of volume ($\sim d^3$). Furthermore, our model offers a clear explanation for the physical origin of the reduced H-trapping ability of GBs compared to relatively open vacancies, as the attractive term of interatomic coupling dominates the H solution in VCs and the repulsive term plays a more significant role at GBs. This elucidates the physical origin of the experimental observation that GBs exhibit weaker H trapping than vacancies in W (*2*, *45*, *46*).

**H solution at the defects of α-Fe and Mo**

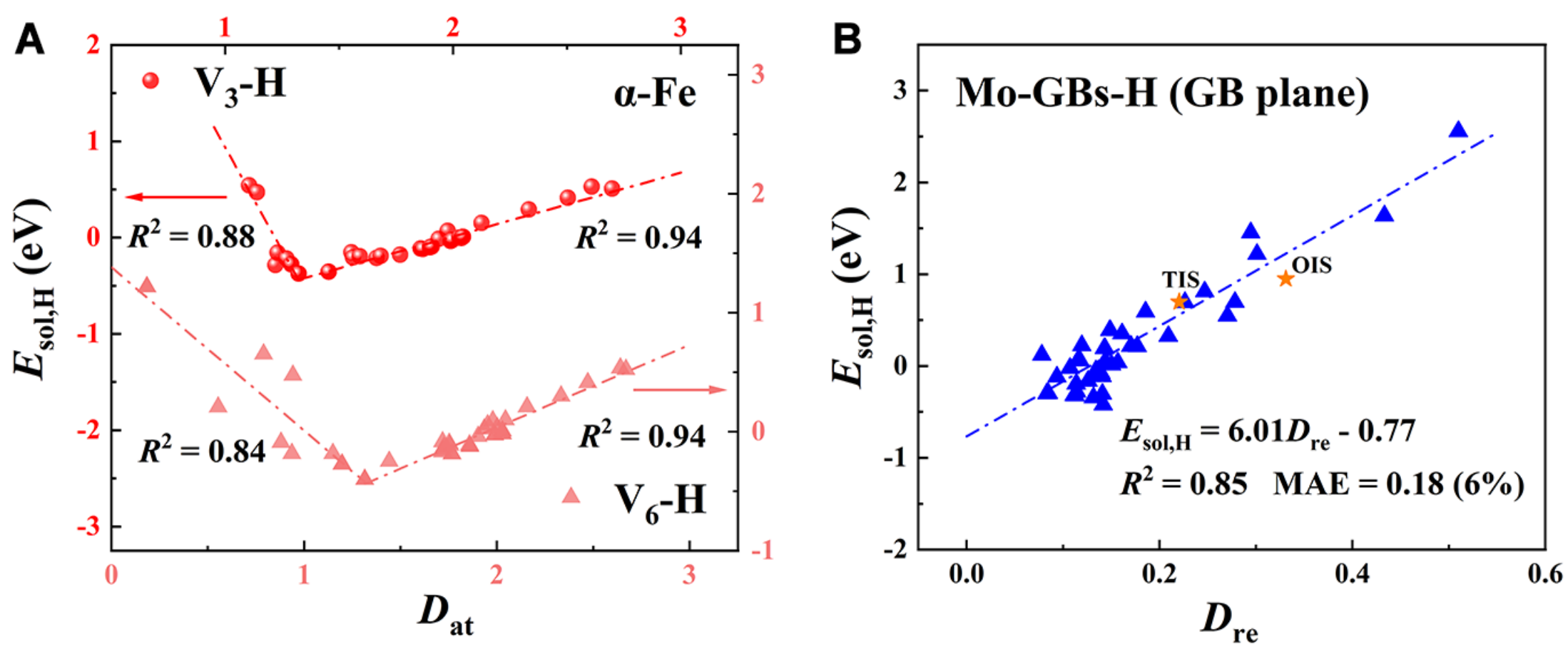

**Fig. 5. The H solution energy ($E_{\mathrm{sol,H}}$) at the VCs of α-Fe and GBs of Mo.** (**A**) The $E_{\mathrm{sol,H}}$ under fully fixed calculations in $V_3$ and $V_6$ of α-Fe versus the descriptor $D_{\mathrm{at}}$. (**B**) The $E_{\mathrm{sol,H}}$ at Mo GB planes versus our descriptor $D_{\mathrm{re}}$.

Our model is also applicable to the H solution energetic across various defects in different bcc metals. For example, our model effectively quantifies the $E_{\mathrm{sol,H}}$ in VCs of α-Fe (Fig. 5A), where the descriptor $D_{\mathrm{at}}$ exhibits a broken-line relationship with $E_{\mathrm{sol,H}}$, analogous to that in W. This broken-line behavior corresponds to H sites transitioning from the VC center to the inner surface and subsequently to subsurface regions, indicating that in different bcc metals, the H solution energetics at different sites in VCs follow a unified rule and share a similar potential-energy landscape. Furthermore, the general applicability of our model is validated by its successful extension to the H solution at molybdenum (Mo) GBs (Fig. 5B). For H in Mo GBs, the repulsive term of the H–M orbital coupling plays a more prominent role and the interactions exhibit localized characteristics, consistent with the behavior in W, demonstrating that the same physical mechanism governs H solution at GBs across different bcc metals. These results demonstrate that our orbital-coupling model is predictive across the different regions of various defects in different bcc metals, underscoring its broad applicability and transferability.

## DISCUSSION

### Understanding of H solution at different defects

Derived from the classical TB model, we build a predictive model from the view of orbital coupling, which provides a deep understanding of the advancements and limitations of the previous empirical models. The previous studies proposed the different empirical arguments depending on the solution sites, local coordination environments and defect types, such as the discrete energy levels and the *CN* of adjacent metal atoms for H at nanovoid inner surfaces (*20*, *24*, *25*), the stress effects for H solution at dislocation loops (*26*, *27*), the available-volume evaluation and polyhedral-deformation models for GBs (*29*, *30*), and so on. There was also a generally accepted viewpoint that the H–H interactions complied with a distance dependence $1/d_{\mathrm{H-H}}^{5}$. However, these different arguments lack a universal theoretical framework and are inadequate for accurately predicting $E_{\mathrm{sol,H}}$ across defects. Notably, our model demonstrates that by combining the three parameters (*d*, *CN*, *n*) into a single descriptor can we establish a universal physical model that achieves a general quantitative prediction of H solution at VCs, vacancy loops and GBs. Compared to the previous descriptors based on empirically geometric structural parameters, our model reflects the underlying electronic-origin mechanism: the $E_{\mathrm{sol,H}}$ at defects strongly depends on the d-band width of neighboring metal atoms and can be quantified with the s–d coupling of H and bcc-metal atoms. This unusual s–d coupling exhibits an anomalous coordination-dependent law for the H–M interactions and the same distance-decay law for both the H–M and H–H interactions ($1/d^{7/2}$). These unexpected results essentially stem from the confined environments of defects, while the claim of $1/d_{\mathrm{H-H}}^{5}$ was misled by the less confined reference interactions, thereby redefining the understanding of H solution at metal defects. Our model is a new development of the classical TB model, establishing a unified theoretical framework for understanding the interatomic bonding at confined environments.

### Application of our model

To validate the across-scale predictive capability of our orbital-coupling model, we performed OKMC simulations of H thermal desorption from heavy-ion-damaged W. Crucially, the H–vacancy binding energies, defined as the difference in the $E_{\mathrm{sol,H}}$ of H at VCs and TIS, are directly predicted by our model and agree well with the DFT results (fig. S7). The de-trapping barriers were then obtained by adding the interstitial H migration energy in bulk W. As shown in Fig. 6A, the simulated TDS curve reproduces the experimental H retention and bimodal profile with remarkable fidelity across the entire heating temperature range (*47*), capturing both the low-temperature peak near 550 K and the high-temperature peak near 800 K.

The origin of the bimodal structure is encoded in the H-concentration dependence of the orbital-coupling model and strongly depends on the distance-decay rule of the H–H interactions. Both the experiment and simulation employ H implantation at 370 K, where all VCs equilibrate at H/V = 2–3 (Fig. 6B). Based on the H–vacancy binding energy predicted by our model as well as the Polanyi–Wigner equation (*20*), we derive the desorption temperatures at different H concentrations (our descriptors at different H concentrations are correlated with the H/V ratio) in Fig. 6C. It shows that the H desorption temperature in VCs decreases as the H/V ratio increases and the desorption temperatures of most clusters converge to the 400–500 K range. Consequently, H is released synchronously from the entire spectrum of VCs as the temperature exceeds ~500 K, producing the first desorption peak near 550 K, where $V_1$ dominates the microstructure (Fig. 6D). The higher-temperature peak, by contrast, is a kinetic emergent phenomenon that is correlated with the VC aggregation and H re-trapping. Once the temperature is high enough for $V_1$ to overcome the migration barrier (1.66 eV (*48*)), VC aggregation redistributes the microstructure. By 700 K, the fraction of $V_1$ plummets from 85.7% to 29.2%, while $V_{5-10}$ and $V_{11-30}$ surge to 42.8% and 23.0%, respectively (Fig. 6D). As VCs grow during annealing, the decrease of the H/V ratio largely eliminates the repulsive H–H interactions, highlighting the dominant role of the H–M interactions according to our model. The resulting stronger H–M binding shifts the effective desorption temperature to the higher value (Fig. 6C). The continuous growth of VCs therefore creates a self-amplified re-trapping cascade: H atoms released from small VCs are captured by freshly formed, larger VCs ($V_{n\geq5}$) whose binding strengths are continuously deepened by modulating the H–H interactions. These results demonstrate the crucial role of the distance-decay rule of H–H interactions in determining the bimodal structure. With 2–3 H atoms per VC, the $1/d_{\mathrm{H-H}}^{5}$ decay generates much shorter-range H–H interactions (Fig. 3D) that produce negligible H–H repulsion in $V_1$, whereas $1/d_{\mathrm{H-H}}^{2}$ does longer-range H–H interactions that yield too strong H–H repulsion in $V_{n\geq5}$ (see table S1). Either of them causes a minor difference of the H–H interactions between $V_1$ and $V_{n\geq5}$ that goes against the formation of bimodal structure. In contrast, only $1/d_{\mathrm{H-H}}^{7/2}$ can generate the significant difference between $V_1$ and $V_{n\geq5}$ for the H–H interactions by ~0.20 eV (and for $E_{\mathrm{sol,H}}$ by > 0.40 eV/H), which is crucial to the formation of the double peaks of H trapping and desorption behavior.

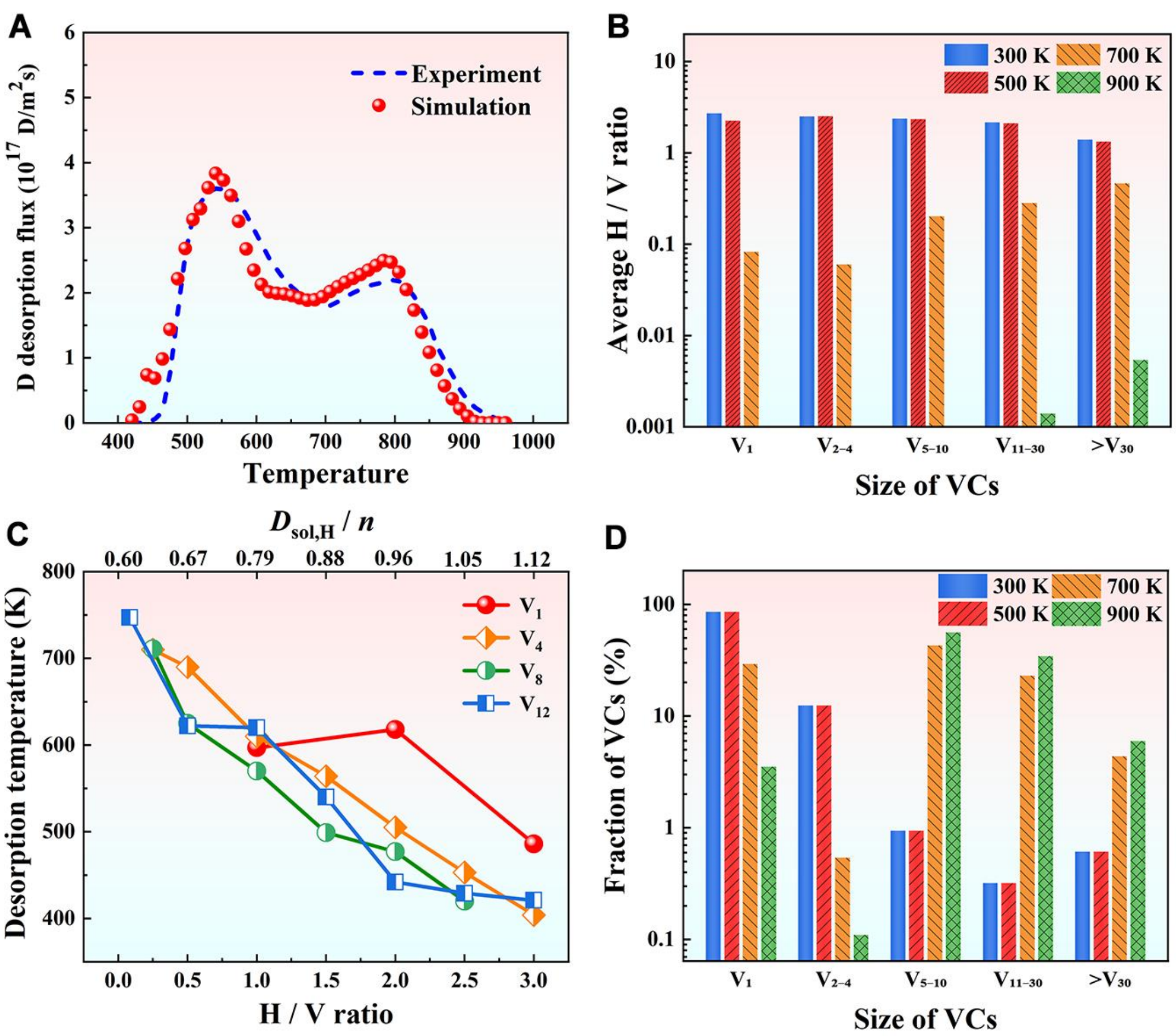

**Fig. 6. Multiscale simulations based on our orbital-coupling model and the OKMC method.** (A) Simulated and experimental TDS profiles. Polycrystalline W samples were firstly irradiated by 20 MeV W ions to 0.23 dpa at 295 K experimentally. After the W irradiation, the sample was exposed to D ions for 74 h at 370 K, then annealed to 1000 K with a heating rate of 3 K/min. (**B**) Average H/V ratio the in VCs of different sizes at 300, 500, 700 and 900 K. (**C**) H desorption temperatures predicted by the orbital-coupling model, relative to the H/V ratio in VCs. $D_{\mathrm{sol,H}}/n$

corresponds to the average of all H–M and H–H orbital couplings at different H/V ratio in $V_4$ (where $n$ is the number of H atoms). (**D**) Evolution of the VC size distribution during annealing.

Overall, the entire bimodal profile is quantitatively reproduced using the H–M and H–H interactions from our orbital-coupling model, providing compelling evidence that the model faithfully encodes the electronic–structure origin of H trapping, transport and release in defective W.

By building an interatomic orbital-coupling rule at metal defects based on coordination number and interatomic distance, we develop a general predictive model to quantify the H solution energetics spanning nanovoids, vacancy loops and GBs in bcc metals. Our model reveals that the confined environments of defects in bcc metals lead to an unconventional s–d coupling rule for both H–M and H–H interactions, generating an anomalous coordination-dependent law for the H–M interactions and forcing the H–H interactions (that usually exhibit the s–s coupling) to follow the distance-decay law of H–M interactions. This model is predictive across the different regions of various defects in different metals and even captures the potential-energy surfaces of H at nanovoids, enabling—when coupled with kinetic simulations—the prediction of the experimental H trapping and desorption properties in bcc metals. Notably, the unusual distance-decay rule of H–H interactions is essential to the experimental bimodal profile of H desorption during TDS experiments. Our predictive scheme reshapes the previous empirical understanding of H–defect interactions, serving as a solid basis for understanding the nature of chemical bonds under constraint and for developing high-performance metallic materials for H-related applications particularly for those with superior resistance to H-induced irradiation damage.

## MATERIALS AND METHODS

### Computational details

All the DFT calculations were performed using the Vienna Ab initio Simulation Package (VASP) code (*49*) with the projector augmented wave (PAW) approach (*50*) and Perdew−Burke−Ernzerhof (PBE) functional (*51*). We used a plane-wave cutoff energy of 400 eV. During the geometric optimization, all atomic positions were fully relaxed within a fixed lattice volume, whereas all atoms remained fixed in the fully fixed calculations. The energies and atomic forces of all calculations were converged within $10^{-5}$ eV and 0.01 eV/Å. We also used a smearing width of 0.1 eV for the Fermi smearing function to facilitate the convergence. The zero-point energy correction was not included due to their approximately unchanged effects (see the comparison with zero-point energy corrections in fig. S8). We employed a fixed-atom scanning method to obtain the potential-energy surface of H in the $V_6$ structure of W. The four-vacancy plane in the $V_6$ of W was selected as the scanning region (Fig. 2A), and a two-dimensional grid of discrete H sites was extracted within this plane, covering the VC centers, the inner surface sites, and the subsurface regions. At each grid point corresponding to an H position, $E_{\mathrm{sol,H}}$ was determined by fixed DFT calculations or our model predictions. After curve smoothing, the potential-energy surfaces for H solution within the four-vacancy plane of the $V_6$ structure were obtained.

For the VCs structures, a 4×4×4 supercell containing 128 atoms was used for calculations, along with a 3×3×3 k-point mesh for Brillouin zone sampling using the Monkhorst–Pack method (*52*). For the GB structures, we constructed 12 symmetric tilt GBs based on the coincidence site lattice (CSL) model (*53*, *54*), including Σ3(110), Σ3(111), Σ3(112), Σ5(210), Σ5(310), Σ7(321), Σ9(411), Σ11(332), Σ13(510), Σ17(410), Σ17(433), and Σ19(611) GBs. The GB structures were constructed using the ATOMSK program (*55*). In order to achieve reasonable GB configurations, we removed one of the overlapped atoms in the initial CSL GB structures. A vacuum layer of 12 Å was applied in the direction of the grain interface to avoid periodic structures. The number of atoms in the GB structures used for calculations, along with the k-point mesh used for Brillouin zone sampling, are presented in table S2.

We employed the GRACE-3L-OMAT-large-ft-AM foundation model (*39*, *40*) to calculate the H–H binding in VCs and H solution at vacancy loops. This foundation model is pretrained on OMat24 (*56*) and fine-tuning on the combined sAlex (*56*) and MPtraj (*57*) datasets. We further fine-tune it using our DFT-calculated energies and forces for H solution at W VCs, optimizing only the reduction weights in the three interaction layers while freeze the remaining weights, and retain the 6 Å cutoff. The energy and force Huber losses (*58*) are assigned weights of 15 and 1, respectively, and training proceeds for four epochs using Adam (*59*) with an initial learning rate of $10^{-4}$ (fig. S4 for the test accuracy ). Using our fine-tuned GRACE potential, we calculated the H–H binding in VCs and relax the vacancy-loop structures in LAMMPS (*60*) using the conjugate-gradient (CG) optimizer until the forces fall below 0.01 eV/Å. The 1/2 <111> vacancy loop of radius 8.30 Å (37 vacancies) is constructed in a 22×22×22 bcc W cell using the ATOMSK program (*55*), with convergence tests confirming the cell size is sufficient for the convergence of the loop formation energy.

The $E_{\text{sol,H}}$ is calculated using the following formula

$$E_{\text{sol,H}} = E_{\text{defect}+n\text{H}} - E_{\text{defect}} - nE_{\text{H}} \quad (4)$$

where $E_{\text{defect}+n\text{H}}$ and $E_{\text{defect}}$ represent the total energy of the VCs, vacancy loop or GBs with and without embedding *n* H atoms, respectively. $E_{\text{H}}$ is the reference energy of H, corresponding to the one-half of the energy of a $H_2$ molecule in vacuum. The H trapping energy, defined as the difference of $E_{\text{sol,H}}$ at defects and TISs, is linearly related to the corresponding $E_{\text{sol,H}}$.

**Multiscale simulations**

Thermal desorption spectra were computed using large-scale OKMC simulations that couple atomistic H–vacancy energetics with evolving irradiation microstructures. General algorithms and parameterizations for defect migration, clustering and dissociation are described in detail in previous work (*25*, *48*). Defect evolution under heavy-ion irradiation was tracked in a 13.02 nm × 13.65 nm × 40.01 nm cell with full periodic boundary conditions, wherein displacement cascades obtained from molecular dynamics were introduced following the experimental PKA spectrum and the NRT damage model to reproduce the depth-dependent damage profile calculated by SRIM (*47*, *61*). H implantation and subsequent thermal desorption were then simulated in a 13.02 nm × 13.65 nm × 2.4 μm supercell with periodic boundaries in the lateral directions and open surfaces normal to the depth axis permitting H outgassing. The de-trapping energy is defined as the sum of the binding energy and the migration energy barrier of interstitial H in W. The H–vacancy binding energy is obtained from our predictive model (see table S3).The migration barriers, attempt frequencies, and $H_2$ recombination and desorption at free surfaces were treated as coverage-dependent activated processes with barriers drawn from the previous work (*25*). Implantation energies, fluxes, fluences and annealing ramps were matched one-to-one with the experimental protocols to enable direct quantitative comparison (*47*).

## Acknowledgments

**Funding:** The authors are thankful for the support from the National Natural Science Foundation of China (Nos. 22173034, 11974128, 12192281), the Jilin Province Natural Science Foundation (20260102094JC), the Fundamental Research Funds for the Central Universities, the computing resources of the High-Performance Computing Center of Jilin University, China.

**Author contributions:** W.G. and H.B.Z. conceived the original idea and designed the strategy. W.G. derived the orbital coupling model with the contribution of Q.Z., with which H.B.Z. and H.X.H derived the kinetic models. Q.Z. performed the DFT calculations, Z.Z. fine-tuned the GRACE potential and performed subsequent calculations, and H.B.Z. and H.X.H performed the OKMC simulations. W.G. and Q.Z. wrote the first draft of the manuscript together, and all the authors jointly discussed and revised the manuscript. Q.Z., H.X.H. and Z.Z drew the figures and prepared the Supplementary Materials with the contribution of W.G. and H.B.Z.. All the authors have discussed and approved the results and conclusions of this article.

**Competing interests:** The authors declare no competing interests.

**Data availability:** All data are available from the corresponding author upon reasonable request.